\documentclass[journal]{journal}
\usepackage{times}
\usepackage{graphicx} 
\usepackage{subfigure} 

\usepackage{algorithm}
\usepackage{algorithmic}

\usepackage{amssymb}
\usepackage{bm}
\usepackage{float}
\usepackage{footmisc}
\usepackage{mathtools}
\usepackage{amsfonts}
\usepackage{amssymb}
\usepackage{caption}

\usepackage{hyperref}

\usepackage[T1]{fontenc}
\usepackage[utf8]{inputenc}
\usepackage{authblk}

\ifCLASSINFOpdf
\else
\fi
\begin{document}
%
\title{Leveraging Sparsity to Speed Up Polynomial Feature Expansions of CSR Matrices Using $K$-Simplex Numbers}
%
%
%

\author[1]{Andrew Nystrom\thanks{nystrom@google.com}}
\author[2]{John Hughes\thanks{jfh@cs.brown.edu}}
\affil[1]{Google AI}
\affil[2]{Brown University}



%
%

\markboth{Journal of \LaTeX\ Class Files,~Vol.~6, No.~1, January~2007}%
{Shell \MakeLowercase{\textit{et al.}}: Bare Demo of IEEEtran.cls for Journals}
%



\maketitle
\thispagestyle{empty}

\begin{abstract}
An algorithm is provided for performing polynomial feature expansions that both operates on and produces compressed sparse row (CSR) matrices.
Previously, no such algorithm existed, and performing polynomial expansions on CSR matrices required an intermediate densification step.
The algorithm performs a $K$-degree expansion by using a bijective function involving $K$-simplex numbers of column indices in the original matrix to column indices in the expanded matrix.
Not only is space saved by operating in CSR format, but the bijective function allows for only the nonzero elements to be iterated over and multiplied together during the expansion, greatly improving average time complexity.
For a vector of dimensionality $D$ and density $0 \le d \le 1$, the algorithm has average time complexity $\Theta(d^KD^K)$ where $K$ is the polynomial-feature order; this  is an improvement by a factor $d^K$ over the standard method.
This work derives the required function for the cases of $K=2$ and $K=3$ and shows its use in the $K=2$ algorithm.
\end{abstract}


%
\IEEEpeerreviewmaketitle

\section{Introduction}

In machine learning and statistical modeling, feature mappings are intra-instance transformations, usually denoted by $x \mapsto \phi(\vec{x})$, that map instance vectors to higher dimensional spaces in which they are more linearly separable, allowing linear models to capture nonlinearities \cite{yuan2012recent}.

A well known and widely used feature mapping is the \emph{polynomial expansion}, which produces a new feature for each degree-$k$ monomial in the original features.
If the original features are $x,y,z$, the order-2 polynomial features are $x^2, y^2, z^2, xy, xz, yz$, and the order-3 polynomial features are $x^3, y^3, z^3, x^2y, x^2z, xy^2, y^2z, xz^2, yz^2, xyz$.
A $K$-order polynomial feature expansion of the feature space allows a linear model to learn polynomial relationships between dependent and independent variables.
This mapping was first utilized in a published experiment by Joseph Diaz Gergonne in 1815 \cite{gergonne1974application, smith1918standard}.

While other methods for capturing nonlinearities have been developed, such as kernels (the direct offspring of feature mappings), trees, generative models, and neural networks, feature mappings are still a popular tool \cite{barker200114, chang2010training, shaw2006intellectual}.
The instance-independent nature of feature mappings allows them to pair well with linear parameter estimation techniques such as stochastic gradient descent, making them a candidate for certain types of large-scale machine learning problems when $D \ll N$.

The compressed sparse row (CSR) matrix format \cite{saad1994sparskit} is widely used \cite{liu2012sparse, bulucc2009parallel, bell2008efficient, white1997improving} and supported \cite{eigenweb, bastien2012theano, scikit-learn, koenker2003sparsem}, and is considered the standard data structure for sparse, computationally heavy tasks.
However, polynomial feature expansions cannot be performed directly on CSR matrices, nor on any sparse matrix format, without intermediate densification steps.
This densification not only adds overhead, but wastefully computes combinations of features that have a product of zero, which are then discarded during conversion into a sparse format.

We describe an algorithm that takes a CSR matrix as input and produces a CSR matrix for its degree-$K$ polynomial feature expansion with no intermediate densification.
The algorithm leverages the CSR format to only compute products of features that result in nonzero values.
This exploits the sparsity of the data to achieve an improved time complexity of $\Theta(d^KD^K)$ on each vector of the matrix where $K$ is the degree of the expansion, $D$ is the dimensionality, and $d$ is the density.
The standard algorithm has time complexity $\Theta(D^K)$.
Since $0 \le d \le 1$, our algorithm is a significant improvement for small $d$.
While the algorithm we describe uses CSR matrices, it can be readily adapted to other sparse formats.

\section{Notation}
We denote matrices by uppercase bold letters thus: $\bm{A}$. 
The $i^{th}$ the row of $\bm{A}$ is written $\bm{a}_i$. All vectors are written in bold, and  $\bm{a}$, with no subscript, is a vector. Non-bold letters are scalars. We sometimes use `slice' notation on subscripts, so that $x_{2:5}$ indicates the second through fifth elements of the vector $x$.

A CSR matrix representation of an $r$-row matrix $\bm{A}$ consists of three vectors: $\bm{c}$, $\bm{d}$, and $\bm{p}$ and a single number: the number $N$ of columns of $\bm{A}$. The vectors
$\bm{c}$ and $\bm{d}$ contain the same number of elements, and hold the column indices and data values, respectively, of all nonzero elements of $\bm{A}$.
The vector $\bm{p}$ has $r$ entries. The values in $\bm{p}$ index both $\bm{c}$ and $\bm{d}$. The $i$th entry $\bm{p}_i$ of $\bm{p}$ tells where
the data describing nonzero columns of $\bm{a}_i$ are within the other two vectors: $\bm{c}_{\bm{p}_i:\bm{p}_{i+1}}$ contain the column indices of those entries; $\bm{d}_{\bm{p}_i:\bm{p}_{i+1}}$ contain the entries themselves.
Since only nonzero elements of each row are held, the overall number $N$ of columns of $\bm{A}$  must also be stored, since it cannot be derived
from the other data.

Scalars, vectors, and matrices are often decorated with a superscript $k$, which is not to be interpreted as an exponent, but instead as an indicator of polynomial expansion: 
For example, if the CSR for $\bm{A}$ is $\bm{c}, \bm{d}, \bm{p}$, then $\bm{c}^2$ is the vector that holds columns for nonzero values in $\bm{A}$'s quadratic feature expansion CSR representation.

Uppercase $K$ refers to the degree of a polynomial or interaction expansion.
When a superscript $\kappa$ (kappa) appears, it indicates that the element below it is in a polynomially expanded context of degree $K$.
For example, if $nnz_i$ is the number of nonezero elements in the vector $i$th row vector of some matrix, $nnz_i^\kappa$ is the number of nonzero elements in the polynomial expansion of that row vector.
Lowercase $k$ refers to a column index.

\section{Motivation}
In this section we present a strawman algorithm for computing polynomial feature expansions on dense matrices.
We then modify the algorithm  slightly to operate on a CSR matrix, to expose its infeasibility in that context.
We then show how the algorithm would be feasible with a bijective mapping from $k$-tuples of column indicies in the input matrix to column indices in the polynomial expansion matrix, which we then derive in the following section.

It should be noted that in practice, as well as in our code and experiments, expansions for degrees $1, 2, \dots, k-1$ are also generated.
The final design matrix is the augmentation of all such expansions.
However, the algorithm descriptions in this paper omit these steps as they would become unnecessarily visually and notationally cumbersome.
Extending them to include all degrees less than $K$ is trivial and does not affect the complexity of the algorithm as the terms that involve $K$ dominate.

\subsection{Dense Second Degree Polynomial Expansion Algorithm}
A natural way to calculate polynomial features for a matrix $\bm{A}$ is to walk down its rows and, for each row, take products of all $K$-combinations of elements.
To determine in which column of $\bm{A}^\kappa_i$ products of elements in $\bm{A}_i$ belong, a simple counter can be set to zero for each row of $\bm{A}$ and incremented after each polynomial feature is generated.
This counter gives the column of $\bm{A}^\kappa_i$ into which each expansion feature belongs.
This is shown in Algorithm \ref{alg:Dense-Second-Order-Polynomial-Expansion}.


\begin{algorithm}
   \caption{Dense Second Order Polynomial Expansion}
   \label{alg:Dense-Second-Order-Polynomial-Expansion}
\begin{algorithmic}[1]
   \STATE {\bfseries Input:} data $\bm{A}$, size $N \times D$
   \STATE $\bm{A}^\kappa$ $\gets$ empty $N \times \binom{D}{2}$ matrix
   \FOR{$i \gets 0$ {\bfseries to} $N-1$}
      \STATE $c_p \gets 0$
      \FOR{$j_1 \gets 0$ {\bfseries to} $D-1$}
          \FOR{$j_2 \gets j_1$ {\bfseries to} $D-1$}
              \STATE $\bm{A}^\kappa_{i{c_p}} \gets \bm{A}_{ij_1} \cdot \bm{A}_{ij_2}$
              \STATE $c_p \gets c_p + 1$
          \ENDFOR
      \ENDFOR
   \ENDFOR
\end{algorithmic}
\end{algorithm}


\subsection{Incomplete Second Degree CSR Polynomial Expansion Algorithm}
\label{sec:final-algo}
Now consider how this algorithm might be modified to accept a CSR matrix.
Instead of walking directly down rows of $\bm{A}$, we will walk down sections of $\bm{c}$ and $\bm{d}$ partitioned by $\bm{p}$, and instead of inserting polynomial features into $\bm{A}^\kappa$, we will insert column numbers into $\bm{c}^\kappa$ and data elements into $\bm{d}^\kappa$.
Throughout the algorithm, we use variables named $nnz$, with sub- or superscripts, to indicate the number of nonzero entries in either a matrix or a row of a matrix. 
See Algorithm \ref{alg:Incomplete-Sparse-Second-Order-Polynomial-Expansion}.


\begin{algorithm}
   \caption{Incomplete Sparse Second Order Polynomial Expansion}
   \label{alg:Incomplete-Sparse-Second-Order-Polynomial-Expansion}
\begin{algorithmic}[1]
   \STATE {\bfseries Input:} data $\bm{A}$, size $N \times D$
   \STATE $\bm{p}^\kappa$ $\gets$ vector of size $N+1$
   \STATE $\bm{p}^\kappa_0 \gets 0$
   \STATE $nnz^\kappa \gets 0$
   \FOR{$i \gets 0$ {\bfseries to} $N-1$}
      \STATE $i_{start} \gets \bm{p}_i$
      \STATE $i_{stop} \gets \bm{p}_{i+1}$
      \STATE $\bm{c}_i \gets \bm{c}_{i_{start}:i_{stop}}$
      \STATE $nnz^\kappa_i \gets \binom{|\bm{c}_i|}{2}$ \label{li:row_nnz_count}
      \STATE $nnz^\kappa \gets nnz^\kappa + nnz^\kappa_i$
      \STATE $\bm{p}^\kappa_{i+1} \gets \bm{p}^\kappa_i + nnz^\kappa_i$
  \ENDFOR
  
  \STATE $\bm{p}^\kappa$ $\gets$ vector of size $N+1$
  \STATE $\bm{c}^\kappa$ $\gets$ vector of size $nnz^\kappa$
  \STATE $\bm{d}^\kappa$ $\gets$ vector of size $nnz^\kappa$
  \STATE $n \gets 0$
  
  \FOR {$i \gets 0$ {\bfseries to} $N-1$}
      \STATE $i_{start} \gets \bm{p}_i$
      \STATE $i_{stop} \gets \bm{p}_{i+1}$
      \STATE $\bm{c}_i \gets \bm{c}_{i_{start}:i_{stop}}$
      \STATE $\bm{d}_i \gets \bm{d}_{i_{start}:i_{stop}}$
      \FOR {$c_1 \gets 0$ {\bfseries to} $|\bm{c}_i|-1$}
          \FOR {$c_2 \gets c_1$ {\bfseries to} $|\bm{c}_i|-1$}
              \STATE $\bm{d}^\kappa_{n} \gets \bm{d}_{c_0} \cdot \bm{d}_{c_1}$
              \STATE $\bm{c}^\kappa_{n} = ?$ \label{li:set_ck}
              \STATE $n \gets n + 1$
          \ENDFOR
      \ENDFOR
  \ENDFOR
\end{algorithmic}
\end{algorithm}

The crux of the problem is at line 25 of Algorithm \ref{alg:Incomplete-Sparse-Second-Order-Polynomial-Expansion}.
Given the arbitrary columns involved in a polynomial feature of $\bm{A}_i$, we need to determine the corresponding column of $\bm{A}^\kappa_i$.
We cannot simply reset a counter for each row as we did in the dense algorithm,  because only columns corresponding to nonzero values are stored.
Any time a column that would have held a zero value is implicitly skipped, the counter would err.

To develop a general algorithm, we require a mapping from a list of $K$ columns of $\bm{A}$ to a single column of $\bm{A}^\kappa$.
If there are $D$ columns of $\bm{A}$ and $\binom{D}{K}$ columns of $\bm{A}^\kappa$, this can be accomplished by a bijective mapping of the following form:

\begin{equation}
(j_0, j_1, \dots, j_{K-1}) \rightarrowtail \hspace{-1.9ex} \twoheadrightarrow p_{j_0j_1 \dots i_{K-1}} \in \{0,1,\dots,\binom{D}{K}-1\} 
\end{equation}

where (i) $ 0 \le j_0 \le j_1 \le \dots \le j_{K-1} < D$, (ii) $(j_0, j_1, \dots, j_{K-1})$ are elements of $\bm{c}$, and (iii) $p_{j_0j_1 \dots i_{K-1}}$ is an element of $\bm{c}^\kappa$. (For interaction features, the constraint is $ 0 \le j_0 < j_1 < \dots < j_{K-1} < D$.)

Stated more verbosely, we require a bijective mapping from tuples consisting of column indicies of the original input to where the column index of the corresponding product of features in the polynomial expansion.
While any bijective mapping would suffice, a common order in which to produce polynomial features is $(0, 1)$, $(0, 2)$, $\dots$, $(0, D-1)$, $(1, 1)$, $(1, 2)$, $\dots$, $(1, D-1)$, $\dots$, $(D-1, D-1)$ for $K=2$ where the elements of the tuples are column indices.
That is, the further to the right an index is, the sooner it is incremented.
If we want for our algorithm to be backwards compatible with existing models, the mapping must use the this same ordering.


\section{Construction of Mappings}

Within this section, $i$, $j$, and $k$ denote column indices.
We will construct mappings for second ($K=2$) and third ($K=3$) degree interaction and polynomial expansions.
To accomplish this, we will require the triangle and tetrahedral numbers.
We denote the $n$th triangle number as $T_2(n) = \frac{n(n+1)}{2}$ and the $n$th tetrahedral number as $T_3(n) = \frac{n(n+1)(n+2)}{6}$.

For reference, we list the first five triangle and tetrahedral numbers in the following table:
\begin{table}
  \centering
  \caption{The first five triangle ($T_2$) and tetrahedral ($T_3$) numbers.}
  \begin{tabular}{| c | c | c |}
    \hline
    $n$ & $T_2(n)$ & $T_3(n)$ \\
    \hline
    0 & 0 & 0 \\
    1 & 1 & 1 \\
    2 & 3 & 4 \\
    3 & 6 & 10 \\
    4 & 10 & 20 \\
    \hline  
  \end{tabular}
\end{table}

\subsection{Second Degree Interaction Mapping}
For second order interaction features, we require a bijective function that maps the elements of the ordered set
\begin{equation}
((0, 1), (0, 2), \dots, (1, 2), (1, 3), \dots, (D-2, D-1))
\end{equation}
to the elements of the ordered set
\begin{equation}
(0,1,\dots,\binom{D-1}{2}-1)
\end{equation}

For $D=4$, we can view the desired mapping $f$ as one that maps the coordinates of matrix cells to $0, 1, 2, 3$.
If we fill the cells of the matrix with the codomain, the target matrix is as follows:
\begin{align}
\begin{bmatrix}
x & 0 & 1 & 2 \\
x & x & 3 & 4 \\
x & x & x & 5 \\
x & x & x & x
\end{bmatrix}
\label{eq:4by4mat_inter}
\end{align}
where the entry in row $i$, column $j$, displays the value of $f(i, j)$.

It will be simpler to instead construct a preliminary mapping, $r(i, j)$ of the following form:
\begin{align}
\begin{bmatrix}
x & 6 & 5 & 4 \\
x & x & 3 & 2 \\
x & x & x & 1 \\
x & x & x & x
\end{bmatrix}
\label{eq:preliminary4by4}
\end{align}
and then subtract the preliminary mapping from the total number of elements in the codomain to create the final mapping.
Note that in equation \ref{eq:preliminary4by4} we have the following relation:

\begin{equation}
T_2(D-i-2) < e^i \le T_2(D-i-1)
\end{equation}

where $e^i$ is the value of any cell in row $i$ of equation \ref{eq:preliminary4by4}.

Therefore, the following mapping will produce equation \ref{eq:preliminary4by4}:

\begin{equation}
r(i, j) = T_2(D-i-1) - (j - i - 1)
\end{equation}

We can now use this result to construct the desired mapping by subtracting it from the size of the codomain:

\begin{equation}
f(i, j) = T_2(D-1) - [T_2(D-i-1) - (j - i - 1)]
\end{equation}

\subsection{Second Degree Polynomial Mapping}
In this case, the target matrix is of the form

\begin{align}
\begin{bmatrix}
0 & 1 & 2 & 3 \\
x & 4 & 5 & 6 \\
x & x & 7 & 8 \\
x & x & x & 9
\end{bmatrix}
\label{eq:4by4mat_poly}
\end{align}

A very similar analysis can be done for the $K=2$ case to yield

\begin{equation}
f(i, j) = T_2(D) - [T_2(D-i) - (j - i)]
\end{equation}

\subsection{Third Degree Interaction Mapping}
For $K=3$ we can no longer view the necessary function as mapping matrix coordinates to cell values; rather, $DxDxD$ tensor coordinates to cell values.
For simplicity, we will instead list the column index tuples and their necessary mappings in a table.
We shall consider the case of $D=5$.

Again, it is simpler to find a mapping $r(i, j, k)$ that maps to the reverse the target indices (plus one) and create a final mapping by subtracting that mapping from the number of elements in the codomain.
We therefore seek a preliminary mapping of the form

\begin{table}
  \caption{Form of the required mapping for $K=3$, $D=5$.}
  \begin{tabular}{| c | c | c |}
    \hline
    $(i, k, j)$ & $r(i,j,k)$ & $f(i,j,k)$ \\
    \hline
    $(0, 1, 2)$ & 10 & 0 \\
    $(0, 1, 3)$ & 9 & 1 \\
    $(0, 1, 4)$ & 8 & 2 \\
    $(0, 2, 3)$ & 7 & 3 \\
    $(0, 2, 4)$ & 6 & 4 \\
    $(0, 3, 4)$ & 5 & 5 \\
    \hline
    $(1, 2, 3)$ & 4 & 6 \\
    $(1, 2, 4)$ & 3 & 7 \\
    $(1, 3, 4)$ & 2 & 8 \\
    \hline
    $(2, 3, 4)$ & 1 & 9 \\
    \hline  
  \end{tabular}
\end{table}

The mapping has been partitioned according to the $i$ dimension.
Note that within each partition is a mapping very similar to the $K=2$ equivalent, but with the indices shifted by a function of $T_3$.
For example, when $i=0$, the indices are shifted by $T_3(2)$, when $i=1$, the shift is $T_3(1)$, and finally, when $i=2$, the shift is $T_3(0)$.
The preliminary mapping is therefore
\begin{equation}
r(i, j, k) = T_3(D-i-3) + T_2(D-j-1) - (k-j-1)
\end{equation}

and the final mapping is therefore

\begin{align}
f(i, j, k) = T_3(D-2) - [&T_3(D-i-3) + \\
                         &T_2(D-j-1) - \\
                         &(k-j-1)]
\end{align}

\subsection{Third Degree Polynomial Mapping}
The analysis of the $K=3$ polynomial case is very similar to that of the $K=3$ interaction case.
However, the target mapping now includes the edge of the simplex, as it included the diagonal of the matrix in the $K=2$ polynomial case.
The analysis yields the mapping

\begin{equation}
f(i, j, k) = T_3(D) - [T_3(D-i-1) + T_2(D-j) - (k-j)]
\end{equation}

\section{Higher Order Mappings}
It can be seen that mappings to higher orders can be constructed inductively.
A $K$ degree mapping is a function of the $K$-simplex numbers and reparameterized versions of all lower order mappings.
However, in practice, higher degrees are not often used as the dimensionality of the expanded vectors becomes prohibitively large.
A fourth degree polynomial expansion of a $D=1000$ vector would have $\binom{1000}{4} = 41,417,124,750$ dimensions.

\section{Final CSR Polynomial Expansion Algorithm}
With the mapping from columns of $\bm{A}$ to a column of $\bm{A}^\kappa$, we can now write the final form of the innermost loop of the algorithm from \ref{sec:final-algo}.
Let the polynomial mapping for $K=2$ be denoted $h^2$.
Then the innermost loop can be completed as follows: 

\begin{algorithm}[H]
   \caption*{Completed Inner Loop of Algorithm \ref{alg:Incomplete-Sparse-Second-Order-Polynomial-Expansion}}
   \label{alg:Inner-Loop-of-Completed-Sparse-Second-Order-Polynomial-Expansion}
\begin{algorithmic}[1]
  \FOR {$c_2 \gets c_1$ {\bfseries to} $|\bm{c}_i|-1$}
      \STATE $j_0 \gets \bm{c}_{c_0}$
      \STATE $j_1 \gets \bm{c}_{c_1}$
      \STATE $c_p \gets h^2(j_0, j_1)$
      \STATE $\bm{d}^\kappa_{n} \gets \bm{d}_{c_0} \cdot \bm{d}_{c_1}$
      \STATE $\bm{c}^\kappa_{n} = c_p$
      \STATE $n \gets n + 1$
  \ENDFOR
\end{algorithmic}
\end{algorithm}

The algorithm can be generalized to higher degrees by simply adding more nested loops, using higher order mappings, modifying the output dimensionality, and adjusting the counting of nonzero polynomial features in line \ref{li:row_nnz_count}.

\section{Higher Degree and Interaction Algorithms}
Most of the steps for higher degrees and interaction expansions (as opposed to polynomial) are the same as for the $K=2$ polynomial case.
The differences are that for higher degrees, an extra loop is needed to iterate over another column index, and a different mapping is required.
For interaction expansions, the column indices are never allowed to equal each other, so each loop executes one less time, and an interaction mapping is required.
Also, for interaction expansions, the way $nnz$ is computed on line \ref{li:row_nnz_count} of Algorithm \ref{alg:Incomplete-Sparse-Second-Order-Polynomial-Expansion}.
Instead of $\binom{|\bm{c}_i|}{K}$, we have $\binom{|\bm{c}_i|-1}{K}$.

\section{Time Complexity}
\subsection{Analytical}
\label{sec:analytical}

Calculating $K$-degree polynomial features via our method for a vector of dimensionality $D$ and density $d$ requires $T_K(dD)$ products.
The complexity of the algorithm, for fixed $K \ll dD$, is therefore
\begin{align}
&\Theta\left(T_K(dD)\right) =\\
&\Theta\left(dD(dD+1)(dD+2)\dots(dD+K-1)\right) =\\
&\Theta\left(d^KD^K\right)
\end{align}

For a matrix of size $N \times D$, the complexity is therefore $\Theta\left(Nd^KD^K\right)$.
The dense algorithm (Algorithm \ref{alg:Dense-Second-Order-Polynomial-Expansion}) does not leverage sparsity, and its complexity is $\Theta\left(ND^K\right)$.
Since $0 \le d \le 1$, the sparse algorithm scales polynomially with the degree of the polynomial expansion.

\subsection{Empirical Results}
To empirically verify the average time complexity of our algorithm, we implemented both the sparse version and the baseline in the Cython programming language so that results would be directly comparable.
We sought the relationships between runtime and the instance count ($N$), the instance dimensionality ($D$), and the instance density ($d$).

\begin{figure*}[ht!]
\vskip 0.2in
\begin{center}
\centerline{\includegraphics[width=\textwidth]{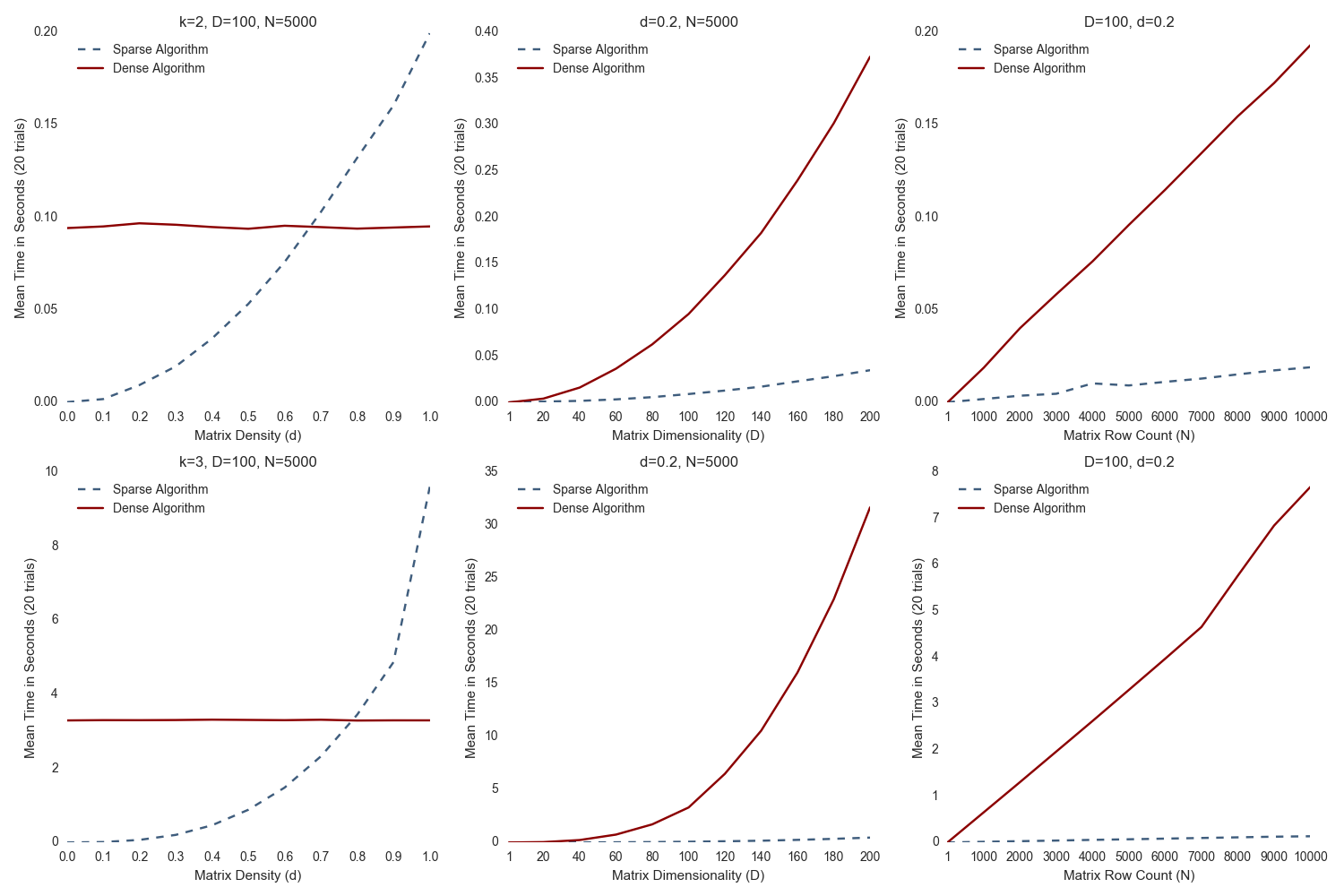}}
\caption{
Summary performance comparison plots for quadratic (top) and cubic (bottom) cases showing how the algorithm's performance varies with $d$, $D$, and $N$; our sparse algorithm is shown in blue, the dense algorithm in red. Each point in each graph is an average of 20 runs, and the time used in densification is not included in the dense-algorithm timings. In the quadratic case, sparsity loses its advantage at about 67\%, and at about 77\% for the cubic case, though these precise intersections depend on $D$. In general, taking advantage of sparsity shows large benefits, so large that it's difficult to see that the performance does not actually change linearly with $D$ (column 2); figure \ref{fig:sparse_D_and_N_vs_time} gives further details.}
\label{fig:all-vs-time}
\end{center}
\vskip -0.2in
\end{figure*}

To find these relationships, we individually varied $N$, $D$, and $d$ while holding the remaining two constant.
For each of these configurations, we generated $20$ matrices and took the average time to reduce variance.
The time to densify did not count against the dense algorithm.
Figure-\ref{fig:all-vs-time} summarizes our findings.

Varying the density ($d$) (column 1) shows that our algorithm scales polynomially with $d$, but that the baseline is unaffected by it.
The runtimes for both algorithms increase polynomially with the dimensionality ($D$), but ours at a significantly reduced rate.
Likewise, both algorithms scale linearly with the instance count ($N$), but ours to a much lesser degree.

Note that the point at which the sparse and dense algorithms intersect when varying $d$ is to the right of $0.5$, which is when a matrix technically becomes sparse.
The point at which this happens will depend on $D$, but the results show that our algorithm is useful even for some dense matrices.

\begin{figure*}[ht!]
\vskip 0.2in
\begin{center}
\centerline{\includegraphics[width=\textwidth]{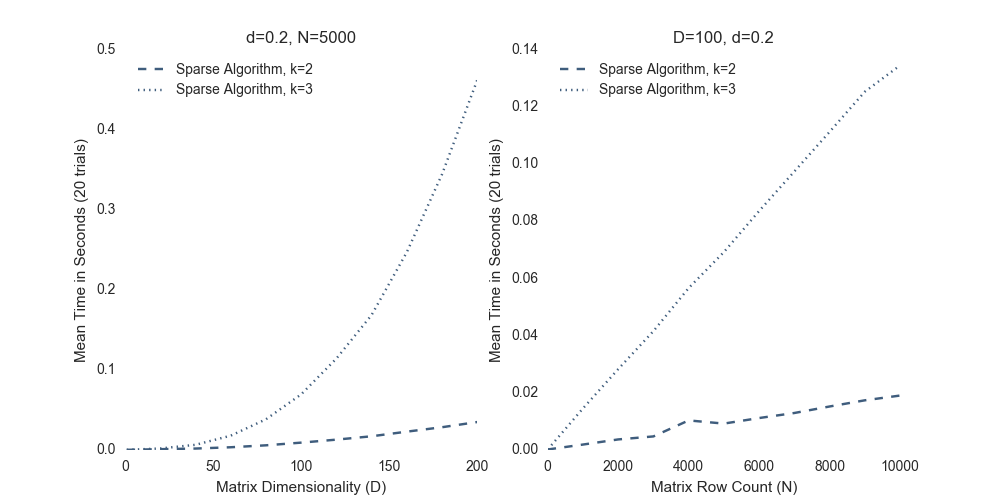}}
\caption{A closer view of only the sparse runtimes while varying $D$ (left) and $N$ (right) for $d = 0.2$. The left subplot shows that varying $D$ gives polynomial growth in runtime; quadratic for $K = 2$ (dashed line) and cubic for $K = 3$ (dotted line). These nonlinearities were not apparent in Figure \ref{fig:all-vs-time} due to the much greater runtimes of the dense algorithm. The right subplot shows linear growth in runtime for both. These findings are in accordance with the analysis of section \ref{sec:analytical}.}
\label{fig:sparse_D_and_N_vs_time}
\end{center}
\vskip -0.2in
\end{figure*}


\section{Conclusion}
We have developed an algorithm for performing polynomial feature expansions on CSR matrices that scales polynomially with respect to the density of the matrix.
The areas within machine learning that this work touches are not en vogue, but they are workhorses of industry, and every improvement in core representations has an impact across a broad range of applications. 
\ifCLASSOPTIONcaptionsoff
  \newpage
\fi



\bibliographystyle{IEEEtran}
%
\bibliography{sparse_poly_proper_order}

%








\end{document}